Utilitarian or Quantile-Welfare Evaluation of Social Welfare?
With Application to Health Cost-Effectiveness Analysis

Charles F. Manski
Department of Economics and Institute for Policy Research, Northwestern University
cfmanski@northwestern.edu

and

John Mullahy
Department of Population Health Sciences, University of Wisconsin-Madison
jmullahy@wisc.edu

November 24, 2026

Abstract

This paper considers quantile-welfare evaluation of social welfare as an alternative to utilitarian evaluation. Maximization of quantile utility has previously been studied as a model of individual decision making under uncertainty, whose properties differ from maximization of expected utility. Particularly noteworthy is that individual maximization of quantile utility requires only an ordinal formalization of utility, not a cardinal one. This paper concerns analysis of social planning rather than individual decision making. We first summarize basic welfare-economic properties of quantile welfare in general terms. We then turn attention to health cost-effectiveness analysis and consider measurement and econometric issues arising in this important applied context. We propose a procedure to nonparametrically bound the quantile welfare of health states using data from binary-choice time-tradeoff (TTO) experiments of the type regularly performed by health economists and provide an empirical example using recent data from the U.K.

Acknowledgements: An earlier version of this paper was circulated with the title “Utilitarian or Quantile-Welfare Evaluation of Health Policy?” We have benefited from the opportunity to present this work in webinars organized by the Center for Health Economics, University of York, and by the EuroQol Research Foundation. Thanks for helpful suggestions and assistance with the data are owed to participants at these webinars and to Bas Janssen, Shaya Nussbaum, Ciaran O'Neill, Julian Reif, Elly Stolk, and Dave Vanness. Our decision to write this paper was stimulated in part by our experiences in the period 2020-2025 as members of the Steering Group (Manski) and Quality-Control Team (Mullahy) advising the EuroQol Foundation on behalf of the UK National Institute for Health Care Excellence (NICE) in its effort to specify a new EQ-5D-5L value set for Health Technology Assessment by the UK National Health Service.

## 1. Introduction

Economists use welfare economics to compare alternative public policies. They assume that an actual or hypothetical planner specifies a social welfare function (SWF). Research seeks to characterize the welfare achieved by alternative feasible policies, aiming to find one that maximizes an SWF. In applications, such research is called cost-benefit analysis by general economists and cost-effectiveness analysis by health economists.

Paul Samuelson placed responsibility for specification of the SWF on society rather than on the economist, writing (Samuelson, 1947, p. 220): "It is a legitimate exercise of economic analysis to examine the consequences of various value judgments, whether or not they are shared by the theorist." In practice, economists have studied policy choice with what might be called *pragmatic* social welfare functions, these being ones motivated by (Manski, 2024, p. 58): "some combination of conjecture regarding societal values, empirical study of population preferences, and concern for analytical tractability."

Economists have mainly studied *personalist* SWFs, ones that are increasing functions of the personal welfare (aka *utility*) of the members of a specified population.[1] They have, moreover, commonly presumed utilitarian aggregation of interpersonally comparable cardinal utilities. In practice, utilitarian analysis implies a focus on population mean utility, which sums individuals' utilities and divides by population size.[2]

Although economists have long found the utilitarian perspective congenial, they have recognized that summation of interpersonally comparable cardinal preferences is only one way to aggregate heterogeneous preferences into a personalist SWF. Rawls (1971) notably argued for a different way, aiming to maximize the utility of the worst-off member of society. Welfare economists have often opined that it is objectionable that utilitarian aggregation is sensitive to cardinal strength of preference. It is sometimes argued that mean preferences may be unduly influenced by individuals with extreme values of cardinal utility.

[1] The term *personalist* SWF revises the word *welfarism* originated by Sen (1977). He wrote (p. 1559): "The general approach of making no use of any information about the social states other than that of personal welfares generated in them may be called 'welfarism.'" Manski (2024) argues that *personalist* SWF expresses Sen's intended distinction more clearly than the word *welfarism*.

[2] While utilitarianism has been the norm, there are exceptions in the extensive literature in welfare and public economics. For example, some research on optimal tax theory has proposed aggregation of a weighted average of cardinal utilities, the weights being determined by observable personal attributes. Saez and Stantcheva (2016) wrote (p. 24): "Weights directly capture society's concerns for fairness without being necessarily tied to individual utilities."

In this paper, we apply quantile-welfare policy evaluation as an alternative to utilitarian evaluation. Manski (1988) proposed and studied maximization of quantile utility as a model of individual decision making under uncertainty, juxtaposing it with maximization of expected utility. The primary motivation was to exploit the fact that maximization of quantile utility requires only an ordinal formalization of utility, not a cardinal one. It was discovered that quantile utility has related interesting properties not shared by maximization of expected utility. A small literature on the subject has developed since then. Rostek (2010) developed an axiomatic interpretation. Manski and Tetenov (2023) studied maximization of quantile utility with sample data.

We transfer these ideas from analysis of individual decision making to analysis of social planning. The substantive context differs. The utility distribution is over heterogeneous states of nature in the individual context and over heterogeneous persons in the social planning context. Nevertheless, the mathematics is similar. Hence, the type of welfare economics used in this paper is easy to explain. Moreover, evaluation of quantile utility is no more difficult to perform than evaluation of expected utility.

Readers with background in social choice theory may recognize that maximization of quantile welfare is essentially the same as *positional dictatorship*, defined in Roberts (1980). As far as we are aware, positional dictatorship has remained an abstract concept in social choice theory, not one studied in applied settings. In addition, many readers familiar with clinical research will already be well acquainted with quantile welfare evaluation criteria although probably not by that name: clinical trials that compare two medical technologies on the basis of median survival time or other time-to-event outcomes are essentially using quantile welfare evaluation.

### 1.1. Utilitarian and Quantile Welfare Evaluation in Health Cost-Effectiveness Analysis

We view the main contributions of this paper to be to applied econometrics and to health economics, not to the theory of welfare economics. We use health cost-effectiveness analysis to flesh out the general welfare economic ideas in a highly prominent and controversial setting. The utilitarian perspective has been especially prevalent in health economics. It has motivated econometric analysis of treatment response to estimate average treatment effects rather than other aspects of distributions of treatment response. In research measuring individual health-related utility in quality-adjusted life years (QALYs), it has motivated calculation of so-called *value sets* for health-related quality of life (HRQoL), which seek to measure the mean valuations of health states in a population.

A subfield of health economics has emerged that uses the EQ-5D framework developed by the EuroQol Group (https://euroqol.org/) for estimating mean HRQoL in a population; see for instance Bauer, Lakdawalla, and Reif (2025). This literature has developed methods to measure mean preferences through surveys that pose hypothetical choice problems and ask respondents to state their preferences among the EQ-5D's multiple health states. See Devlin *et al.* (2022). The EQ-5D framework is hardly known in general economics and moderately familiar in broad health economics, but it is prominent in streams of health economics that focus on evaluation. An especially noteworthy case is its use by the UK National Institute for Health and Care Excellence (NICE) to assess the population health benefits of new medical treatments, as an input to National Health Service decisions to fund new treatments.

Research in health economics has not previously studied quantile-welfare evaluation of new medical treatments. However, the literature has on occasion come close when health economists have suggested the use of median rather than mean population HRQoL to evaluate treatments. Such suggestions have sometimes been motivated empirically by the well-known insensitivity of the median to the particular shape of the tails of probability distributions, a property loosely called invariance to "outliers." They have sometimes been motivated conceptually by the heuristic notion that the median is a better summary statistic of "central tendency" than the mean when a distribution is skewed (see, Austin, 2002, p. 336). The median, of course, is a synonym for the 0.5-quantile. Our motivation is different. We are not mainly concerned about invariance to outliers or to distibutional skewness. We do not single out the median for special attention among all quantiles.

### 1.2. Plan for the Paper

Section 2 summarizes basic theoretical properties of quantile welfare. Section 3 considers the general empirical problem of learning population preferences. Section 4 discusses measurement and empirical considerations arising in the specific context of quantile welfare evaluation of health-related utility, when the data are from binary-choice time-tradeoff (TTO) experiments of the sort often performed in HRQoL applications. Section 5 presents an empirical illustration of such welfare evaluation using recent U.K. EQ-5D-5L data on valuations of health states. Section 6 concludes.

## 2. Maximization of Quantile Welfare

### 2.1. Individual Maximization of Quantile Utility

Manski (1988) proposed maximization of quantile utility as a class of criteria for individual decision making under uncertainty. Both quantile and expected utility maximization respect weak stochastic dominance, a basic feature of any reasonable decision rule. Nevertheless, maximization of expected and quantile utility differ in important respects. The most fundamental is that the ranking of actions by expected utility is invariant only to cardinal transformations of the utility function. In contrast, rankings by quantile utility are invariant to ordinal transformations.

This fundamental difference has important consequences. An immediate technical difference is that expected utility is not well-defined when the distribution of utility has unbounded support with fat tails, but quantile utility is always well-defined. A more subtle substantive consequence is that quantile utility yields a significant generalization of traditional ideas of risk preference.[3]

Our analysis of quantile welfare formalizes a perspective on inequality preference, not risk preference. Yet the math is essentially the same. Hence, we find it useful to summarize the quantile utility perspective on risk preference in the next subsections, as a prelude to our introduction of quantile welfare analysis in Section 2.2.

#### 2.1.1. The Quantile Utility Perspective on Risk Preference

From the quantile utility perspective, it is reasonable to define one action to be riskier than another if the utility distribution function of the latter crosses that of the former from below. Lemma 2 of Manski (1988) shows that this single-crossing property is equivalent to 'spreading' the less risky utility distribution in a natural manner.

The single-crossing criterion for comparing the riskiness of actions is much more general than the accepted characterization of riskiness in expected utility theory. With expected utility, actions are risk comparable only if the single-crossing property and other conditions hold, being: (1) utility is an increasing function of a real-valued outcome, such as income or health, and (2) the actions being compared have outcome distributions with the same mean. Then one action is deemed riskier than another if its outcome distribution is a mean-preserving spread of the other. With quantile utility, the outcomes generating utilities are unrestricted and utility need only be a measurable function of the outcome. Outcome distributions need not have the same means; indeed, the means need not exist.

With the risk comparability of actions defined by the single-crossing property, Proposition 3 of Manski (1988) shows that the risk preference of a quantile utility maximizer increases with the utility distribution quantile that they maximize. The reasoning underlying the Proposition is simplest when considering two actions, say A and B, whose utility distributions are continuous and strictly increasing.

[3] We emphasize that this paper's quantile utility welfare comparisons are those of Manski (1997), termed ΔD evaluations, that compare quantiles of the marginal distributions of utilities that arise under different actions or policies. In contrast, DΔ evaluations compute quantiles of the distribution of the difference in utility under alternative policies. ΔD evaluations are common in clinical research where, for instance, differences in marginal medians of time-to-event outcome distributions are often of primary interest (Mullahy, 2021).

Suppose that the distribution function for A crosses that of B from below at some utility value $u^*$, where the distribution functions both take some value $p^*$. Thus, $P[u(A) \leq u^*] = P[u(B) \leq u^*] = p^*$. Then the α-quantile of A is larger than that of B for all $\alpha < p^*$, equals that of B for $\alpha = p^*$, and is smaller than that of B for all $\alpha > p^*$. Hence, an individual who maximizes the α-quantile of utility strictly prefers A to B if $\alpha < p^*$, is indifferent if $\alpha = p^*$, and strictly prefers B to A if $\alpha > p^*$.

The quantile-utility characterization of risk preference contrasts sharply with that in expected utility theory. There, individuals with concave utility functions are called risk-averse and ones with convex utility functions are called risk loving. Concavity and convexity are cardinal properties of functions, which are not germane in quantile utility theory.

### 2.1.2. Sequential Quantile Utility Maximization

The expected utility criterion has a property that initially appears more appealing than quantile utility criteria. Consider actions A and B such that the utility distribution of A stochastically dominates that of B strictly. Then the expected utility of A is strictly larger than that of B, but some quantile utilities of A may equal those of B. Thus, quantile utility maximization may indicate indifference between actions A and B, even though the utility distribution of A stochastically dominates that of B strictly.

To address this issue, Manski (1988) proposed sequential quantile utility maximization, a lexicographic refinement where the individual establishes a sequence of quantiles for consideration. If one action yields a larger value of the first quantile in the sequence, the individual strictly prefers this action. If both actions yield the same value of the first quantile in the sequence, the individual compares them using the second quantile and strictly prefers one to the other if they differ in the second quantile. If not, the individual considers the third quantile, and so on. Levy and Kroll (1978) proved that one probability distribution stochastically dominates another strictly if and only if all quantiles of the former distribution are weakly larger than those of the latter and at least one quantile is strictly larger. Hence, the result of sequential quantile utility maximization is that strict stochastic dominance yields strict preference.

## 2.2. Maximization of Quantile Social Welfare

Whereas decision theory studies an individual who chooses an action in an unknown state of nature, personalist welfare economics studies a planner who chooses a policy for a population of individuals. This is a profound substantive difference, but the mathematics of the two settings are similar. The population of welfare economics is analogous to the state space of decision theory.

The mathematical analogy strengthens when one places a probability distribution on the state space in decision theory and, correspondingly, characterizes the population in welfare economics by a distribution of utility functions. Then the individual and the planner both associate each feasible action with a probability distribution, respectively a distribution of utility across the state space and one across the population. In both settings, the decision maker may rank actions by some functional of the relevant distribution, such as the mean or a quantile.

For simplicity, we adhere to the usual practice in personalist welfare economics of evaluating the social welfare of deterministic personal outcomes. Thus, we do not consider social welfare functions that incorporate considerations of how individuals make decisions under uncertainty. While an alternative approach combining considerations of population heterogeneity with individual-level state-space uncertainty might yield useful insights beyond those offered by our analysis, such analysis is beyond this paper's scope.

Whereas decision theory has considered respect for stochastic dominance to be a basic feature of any reasonable decision criterion, personalist welfare economics has taken respect for Pareto dominance to be a sine qua non. The latter property is not probabilistic, as it views each member of the population to be a named individual. However, it is usual in welfare economics to suppose that the planner is concerned only with the population distribution of utility, not with the identities of population members. When this condition is imposed, the essence of Pareto dominance is conveyed by stochastic dominance.

Risk preference in decision theory is mathematically analogous to inequality preference in welfare economics. In both cases, the concern is with some measure of the spread of the relevant probability distribution. In utilitarian welfare economics, an inequality-averse planner maximizes the mean of a concave transformation of interpersonally comparable cardinal utility, with strength of inequality aversion being expressed by the degree of concavity. A leading example is optimal income tax theory as initiated by Mirrlees (1971). With quantile-welfare maximization, aversion to inequality is conveyed by the quantile maximized, a lower quantile expressing greater aversion. This is the central idea underpinning value-at-risk metrics in insurance and actuarial research. The limiting case is the SWF proposed by Rawls (1971), which evaluates social welfare by the utility of the worst-off (smallest quantile) member of the population.

## 3. Learning Population Preferences

A canonical policy evaluation setting in the spirit of Section 2 contemplates a social planner who (a) is confronted with policy options (A, B, C, . . . ) that induce welfare distributions ($P_A$, $P_B$, $P_C$, . . .) and (b) strives to implement the best policy from among these options. What is "best" depends on the planner's goals. What is best for a utilitarian planner (say option A) may differ from what is best for a quantile-welfare maximizing planner (say B).

Choosing an optimal policy requires the planner to know the relevant properties of the policy-specific welfare distributions, whether their means or quantiles. The primary practical impediment is empirical. Determination of optimal policy requires knowledge of the distribution of preferences of population members. This requires empirical research that yields knowledge of population preferences.

It is not realistic to expect much progress based on econometric analysis of revealed preferences. The core of the identification problem was understood by Samuelson (1938, 1948) in his pioneering study of revealed preference analysis. Consider a rational person making decisions with full knowledge of utility outcomes. Observation of a choice reveals only that this action is weakly preferred to all other elements of the choice set. It does not reveal the full preference ordering. Modern econometric revealed preference analysis shows that unrealistically strong assumptions are required to point-identify population distributions of utility functions. Study of partial identification with weaker assumptions shows that credible assumptions commonly yield a large identification region for the distribution of utility functions. See Manski (2014, 2024),

We think it more promising to focus research on performance of stated-choice analyses, in which a researcher poses multiple hypothetical choice settings to survey respondents and ask them to predict the choices they would make in each setting. If the respondents are a random sample of the population of interest, the data may be used to estimate the distribution of preferences in the population. Inference on preferences from data on stated choices has a long history in econometric analysis of discrete choice. Ben Akiva, McFadden, and Train (2019) provide a comprehensive review of the conventional practice, which asks persons to state deterministic choices.[4]

Whatever the specifics of data collection and interpretation, a huge advantage of stated-choice analysis over revealed-preference analysis is that the choice settings considered are not limited by what nature offers. A researcher can elicit predictions of behavior in a wide spectrum of hypothetical choice settings.

Health economists have widely used stated-choice data in utilitarian assessment of medical treatments. The prevalent practice has been to measure health-related cardinal utility on a scale called a quality-adjusted life year (QALY). Weinstein, Torrance, and McGuire (68) describe the scale as follows on p.S5:

> "Health states must be valued on a scale where the value of being dead must be 0, because the absence of life is considered to be worth 0 QALYs. By convention, the upper end of the scale is defined as perfect health, with a value of 1. To permit aggregation of QALY changes, the value scale should have interval scale properties such that, for example, a gain from 0.2 to 0.4 is equally valuable as a gain from 0.6 to 0.8."

[4] A caveat is that it may not be credible to analyze stated choices as if they were actual ones. Multiple issues may arise. Manski (1999) cautioned that stated choices may differ from actual ones if researchers provide respondents with different information than they would have when facing actual choice problems. The norm has been to pose *incomplete scenarios*, where respondents are given a subset of the information they would have in actual choice settings. When scenarios are incomplete, stated choices are point predictions of uncertain actual choices. To cope with this, Manski (1999) proposed elicitation of choice probabilities rather than deterministic stated choices and showed how the elicited choice probabilities may be used to estimate random utility models.

Although the basic QALY scale is [0, 1], health economists often consider the possibility that persons may view some adverse health states as having value less than 0, thus extending the scale. We explain further in the next section.

To our knowledge health economists have not used statedchoice data to estimate quantiles of health-related utility. Section 4 shows how this may be accomplished and section 5 presents an empirical illustration.

## 4. Bounding Quantile Welfare in Health Cost-Effectiveness Analysis, Using Interval Data from Binary-Choice Time-Tradeoff Experiments

We now turn to a prominent health policy application of the ideas discussed in sections 2 and 3. When assessing health policy, the data that serve as the basis of welfare evaluations may be cardinal (either point- or interval-measured) or ordinal, and may describe either benefits *per se* or may amalgamate benefits and costs into a single measure (see Stinnett and Mullahy, 1998). Quantile welfare evaluation is well equipped to handle any of these data structures.[5] In this section we consider measurement and evaluation of health-related utility using stated-choice data from binary-choice experiments. Such data have an interval-measurement character, which complicates estimation of preference distributions.

### 4.1. Idealized Estimation of an Abstract Distribution of Health-Related Utility

Henceforth, let H be a time-invariant set of potential health states that could be realized in each year t = 1, . . . , T, where T is a specified terminal year. Each state in year t is a value $h_t \in H$, t = 1, . . . . , T. At each t, one value in H will be realized. Following a convention long used in health economics, we use the normalizations $h_t = 1$ to denote full health and $h_t = 0$ to denote death in a year before or equal to t. All health state vectors $h \equiv (h_t, t = 1, \ldots, T)$ are feasible, except that $h_t = 0 \Rightarrow h_{t+1} = 0$. For now we take $h_t$ as scalar; we relax this assumption later.

Let J denote the population of interest. For each $j \in J$, let $u_j(h)$ denote the person-specific ordinal utility of state vector h to person j. We make no assumptions about the structure of preferences except for two properties, one a technical regularity condition and the other a realistic substantive condition. The technical property is that preferences are almost always strict rather than weak; that is, indifference occurs at most with probability zero across the population. This assumption has been standard in study of random utility models, its advantage being that it makes choice probabilities well-defined. The substantive condition, which has been assumed throughout health economics, is that full health in a given year is preferred to other health states, ceteris paribus. That is, let h be any feasible state vector, and let $h^* = h$ except that it replaces some components of h that do not equal 1 with components that equal 1. Then $u_j(h^*) > u_j(h)$.

The broad definition of health states and utility posed here encompasses the tighter specifications commonly assumed in research on health economics. It has been particularly common to assume that utility has the undiscounted, additive QALY form $u_j(h) = \sum_{t=1}^{T} q_j(h_t)$, where $q_j(h_t)$ is person j's quality of life in health state $h_t$, $q_j(1) = 1$, and $q_j(0) = 0$. Health economists occasionally consider a time-discounted extension of this utility function of the form $u_j(h) = \sum_{t=1}^{T} (\delta_j)^t q_j(h_t)$, where $\delta_j$ is the rate of time discount used by j.

In principle, each $j \in J$ may be presented with binary choice experiments that draw pairs of feasible health state vectors and ask the person to choose between them. With sufficient data collection, this reveals $u_j(\cdot)$. If the entire population is surveyed and all respond, it reveals the utility distribution $P[u(\cdot)]$.

Thus far, utility functions are not interpersonally comparable. Defining quantiles requires a reasonable way to make them ordinally comparable. The maintained assumption that full health in a given year is preferred to other health states, ceteris paribus, provides a way to proceed.

Let d denote year of death and consider the health state vector $h^{*d} \equiv (1, \ldots 1, 0, \ldots 0)$ in which a person is alive with full health through year d and dies at the end of year d. We make ordinal utilities interpersonally comparable by letting $u_j(h^{*d}) = d$ for all $j \in J$ and $d \in (1, \ldots T)$. Similarly, immediate death at the time of the choice experiment has utility $u_j(h^{*0}) = 0$, $j \in J$. These interpersonal normalizations are consistent with the assumption that all members of the population prefer full health to other health states, ceteris paribus. In particular, it is consistent with the undiscounted QALY form of utility.

Now consider the utility $u_j(h)$ of any health state vector h. We can use the $h^{*d}$ normalization to place $u_j(h)$ in one of T + 1 intervals, as follows:

- If TTO experiments reveal that $u_j(h) < u_j(h^{*0})$, then $u_j(h) < 0$.
- If TTO experiments reveal that $u_j(h^{*(d-1)}) < u_j(h) < u_j(h^{*d})$ for d such that $0 < d \leq T$, then $d - 1 < u_j(h) < d$.

It follows that we can, without further assumptions, place the α-quantile of u(h) in one of T + 1 intervals. These are

- $Q_\alpha[u(h)] < 0$ if $P[u(h) < 0] \geq \alpha$.
- $d - 1 < Q_\alpha[u(h)] < d$ if $P[u(h) < d - 1] < \alpha$ and $P[u(h) < d] \geq \alpha$, $0 < d \leq T$.

The above derivation supposes that a survey of the entire population enables precise determination of $P[u(h) < d]$, d = 0, . . . T. In practice, researchers may draw a random sample of the population and assume that nonresponse is random. Then empirical frequencies yield consistent estimates of these probabilities.

### 4.2. Empirical Considerations in QALY Measurement

Whereas the utilities of health state vectors were considered abstractly above, we now specialize to the QALY form of health-related utility prevalent in health economics. This section describes econometric issues faced in empirical research aiming to measure the distribution of QALYs in a population, given a specified medical treatment or other health intervention.[6] Instructive discussions of some issues from the perspective of one regulatory body, the UK National Institute for Health and Care Excellence (NICE), are found in chapter 4 of NICE (2025). After defining basic concepts in Section 4.2.1, we consider estimation of mean QALYs in Section 4.3, paying particular attention to problematic Tobit modeling strategies that are used often in applied evaluations. We then turn attention to quantile welfare evaluation in Section 4.4.

[6] Many considerations important in empirical practice are not discussed here, including the structures of TTO survey elicitations, sampling designs, sampling execution (including interviewer effects), cognitive burdens and challenges for subjects (e.g. understanding TTO logic). While outside the focus of this paper, one such consideration we view as essential to address concerns potential disconnects between conceptual characterizations of better and worse health states (EuroQol Research Foundation, 2023, page 20) and empirical findings that may not adhere to the monotonicity relationships implied by the conceptual characterizations. Post hoc repair of such empirical findings by imposing monotonicity relationships on such disobedient parameter estimates strikes us as problematic.

### 4.2.1. The Anatomy of QALYs

NICE (2025), Chapter 4 describes the nature of a QALY measure in paragraph 4.3.2:

> "4.3.2. A QALY combines both quality of life and life expectancy into a single index. In calculating QALYs, each of the health states experienced within the time horizon of the model is given a utility reflecting the health-related quality of life associated with that health state. The time spent in each health state is multiplied by the utility. Deriving the utility for a particular health state usually comprises 2 elements: measuring health-related quality of life in people who are in the relevant health state and valuing it according to preferences for that health state relative to other states (usually perfect health and death)."

Regarding the time horizon of the experienced health states, NICE (2025) writes:

> "4.2.22. The time horizon for estimating clinical effectiveness and value for money should be long enough to reflect all important differences in costs or outcomes between the technologies being compared."
> "4.2.23. Many technologies have effects on costs and outcomes over a patient's lifetime. In these circumstances, a lifetime time horizon is usually appropriate. A lifetime time horizon is needed when alternative technologies lead to differences in survival or benefits that last for the remainder of a person's life."
> "4.2.25. A time horizon shorter than a patient's lifetime could be justified if there is no differential mortality effect between technologies and the differences in costs and clinical outcomes relate to a relatively short period."

Recalling that $QALY_j(h) = u_j(h) = \sum_{t=1}^{T} q_j(h_t)$, the central considerations in measuring QALYs in practice concern measurement of the $q_j(h_t)$ and determination of the relevant time horizon(s) over which the $q_j(h_t)$ are to be learned. The convention in EQ-5D studies has been to make the horizon ten years.

Incremental QALYs between two treatments 0 and 1 that may yield different health-state time patterns are

(1) $QALY_j(h^1) - QALY_j(h^0) = \sum_{t=1}^{T}[q_j(h_t{}^1) - q_j(h_t{}^0)]$,

where $h^0$ and $h^1$ are the health state vectors experienced under treatments 0 and 1. If time to death varies between treatments, using the same time horizon T for both treatments is not problematic because $q_j(h_t{}^k) = 0$ at any time period at or after death. The recognition that, for a given treatment and individual, health states may vary over time is consistent with the paragraphs from NICE (2025) excerpted above.

In applications, measurement of QALYs typically relies on at least two types of data sources. One (D1) is a clinical (e.g. RCT) or population study that estimates the time patterns of situation-relevant (e.g. treatment-specific) health states $h_t{}^k$. The other (D2) is a population-level elicitation of the utilities associated with the health states.[7]

Much of the empirical QALY literature has focused on estimation of population mean QALYs, E(QALY), with QALY defined as above. One can conceive of E(QALY) using the iterated expectation

(2) $E(QALY) = E[E(u(h)|h)]$.

The inner expectation fixes the health state h and uses the D2 data to learn its mean across the population distribution of $u_j(h)$. The outer expectation uses the D1 data to learn the mean of the inner expectation with respect to the population distribution of health states $h^k = (h_{jt}{}^k$, all j, all t) that would occur under each treatment k.

Quantile welfare evaluation does not use knowledge of E(QALY). Instead, it uses knowledge of the relevant quantile of the population distribution of the QALYs measured using D1 and D2. While there is no corresponding law of iterated quantiles, the idea is similar. That is, variation in measured QALYs across the population arises from distinct two sources. One is individual variation in utility functions and the other is individual variation in the health states generated by a specified treatment.

[7] A third type of data (D3) is sometimes relevant. For instance, a key clinical study may study outcome measures y that differ from the EQ-5D categories h. In this case a data set that allows one to "map" between y and h may be used. See Wailoo et al. (2023).

### 4.3. TTO Measurement of Health State Utilities and Tobit Estimation of Mean Quality of Life

EuroQol Foundation (2023) describes EQ-5D-based health-state utilities as follows:

> "Each health state can potentially be assigned a summary index score based on societal preference weights for the health state. These weights, sometimes referred to as 'utilities', are often used to compute QALYs for use in health economic analyses. Health state index scores generally range from less than 0 (where 0 is the value of a health state equivalent to dead; negative values representing values as worse than dead) to 1 (the value of full health), with higher scores indicating higher health utility. The health state preferences often represent national or regional values and can therefore differ between countries/regions."

Regarding empirical measurement, the range of observed health-state utilities in conventional TTO choice experiments is a bounded interval [L, U], where U is one ("the value of full health") and $\infty < L \leq 0$ ("equivalent to dead" or "worse than dead," but not infinitely worse than dead). See Jakubczyk (2023) for discussion of interpersonal comparability of utilities when worse-than-dead outcomes are possible.

Empirical analysis of QALYs using EQ-5D data has mainly focused on estimation of the mean of $q(h_t)$ in some population of concern. See, for example, the analysis recently performed for the population of the UK (Rowen et al., 2026). Note that if each health-state utility $q(h_t)$ is bounded in [L, U], then each of the t quality of life summands in the summation $\sum_{t=1}^{T} q_j(h_t)$ is also bounded in [L, U], so the total QALYs is bounded in [TL, TU]. With EQ-5D $h_t$ is now a five-dimension vector.

Research using EQ-5D has generally interpreted the upper and lower bounds U and L differently. It has been presumed that health-related utility cannot logically exceed that of "full health," so U = 1 is a true upper bound on utility. However, there is no similar consensus on how low individuals may perceive health states that are “worse than dead.” Hence, the lowest observed utility L is viewed as an artifact of the TTO measurement protocol, yielding a censored value of the true negative utility of very poor health states.

The EQ-5D process using the Tobit model (Tobin, 1957) to estimate value sets has been controversial for econometric, psychological, and normative reasons. The econometric concern is that Tobit model estimates are fragile, being sensitive to departures from normality in the actual distribution of cardinal utilities. The psychological concern is that personal evaluation of their utility in very poor health states is a challenging cognitive task, suggesting that individuals may not have well-defined utilities for such states. The normative concern is that, even if individuals would express very negative values of health-related utility, society may not want to make policy using an SWF that is “unduly influenced” by these expressions.

In health economic analysis using the EQ-5D framework, extremely high values of HRQoL are prevented by the convention of defining the value 1 to express “full health,” which cannot be exceeded by definition. The value 0 is routinely used to express the value of death or a health state equivalent to dead, but individuals are not required to view death as worse than any living state of health. Empirical analysis of stated preference experiments regularly finds that some individuals view certain hypothetical states of health to be “worse than dead,” expressing negative values of HRQoL in these states. The standard EQ-5D survey protocol only permits respondents to express a HRQoL that is moderately worse than death, providing no way to express very negative values. Analysts have interpreted this as generating a data censoring problem. They have commonly dealt with it by estimation of a Tobit model, which assumes that utilities in a specified health state are normally distributed. Again see Devlin *et al*. (2022).

To cope with the assumed censoring, the UK NICE and other agencies have used TTO data to estimate a Tobit model of the inner expectation E(u(h)|h), with estimation by the maximum likelihood method (Amemiya, 1973). Indeed, the study protocol for the UK valuation of EQ-5D-5L initiated in 2020 explicitly committed estimation using versions of the Tobit model. See Rowen *et al*. (2023, 2026). We provide details in the Appendix.

4.4. Quantile Welfare Evaluation of QALYs

To inform quantile welfare evaluation requires estimation of the decision-relevant quantiles of the distribution of the QALYs $\sum_{t=1}^{T} q_j(h_t)$, conditional on relevant covariates x. Nonparametric quantile estimation would be ideal.

Limitations of sample size may make parametric estimation desirable in practice. Suppose one considers observed quality of life q measured in the bound [L, U] to be censored values of latent quality of life in $(-\infty, \infty)$, and that one observes individual covariates x. Then it is natural to consider implementing a doubly-censored quantile regression strategy proposed initially by Powell (1986); see also Koenker (2017). This method assumes that the quantile of interest is a linear function of x, but it makes no other substantive distributional assumption. In sharp contrast to the Tobit model, it does not require one to assume that quality of life is normally distributed and homoskedastic.

Importantly, Powell's approach does not require one to conceive of the values q = L or q = U as arising from a censoring process. In general, one observes non-zero point mass of the empirical distribution of $q^*$ at L and U. If one interprets L as a fixed lower bound on the measurement of utility, then all quantiles of P are point identified. If one interprets L as a left-censoring point, then some quantiles of P—specifically those where the probability mass at L is larger than the quantile(s) under consideration—are not point identified. Being agnostic about the nature of L, one can say that the quantiles of P where the probability mass at L is larger than those quantiles are at most L.

We have performed an illustrative simulation to showcase key features of evaluation using quantile versus utilitarian welfare in populations with heterogeneous preferences and health outcomes. The details are available in an earlier version of this paper at https://arxiv.org/abs/2509.05529v3.

5. Empirical Analysis

This paper's empirical analysis draws on data from a representative sample of the U.K. public used to generate the EQ-5D-5L[8] value set developed by the National Institute for Health and Care Excellence (NICE) for U.K. health technology assessments (see Rowen et al., 2026, and National Institute for Health and Care Excellence, 2025). Our analysis illustrates how evaluation based on quantile welfare criteria can yield inferences and recommendations that may differ from those arising from utilitarian welfare analysis.

Rowen et al. (2026) provide a detailed description of the sample (see also Stolk et al., 2019, for important technical details). In brief, the sample consists of 1,200 subjects, each of whom was asked to engage in composite time-tradeoff (cTTO) experiments for ten pre-assigned EQ-5D-5L health states drawn from a pre-determined set of 102 of the possible 3,125 states. These exercises result in cTTO utility values $u_{jk} = u_k(H_j)$ for subject k and health state j. The final sample consists of 10,680 observations that represent the ten cTTO valuation exercises performed by each of 1,200 respondents after eliminating 1,320 observations for which the cTTO responses were flagged for possible response-quality problems (as was done by Rowen et al.). The cTTO method accommodates health states that may be considered worse than dead (WTD) by some subjects.

Our main interests are in illustrating (a) how quantile and utilitarian welfare criteria may yield substantively different recommendations, and (b) how considerations of different quantiles (here we consider .25, .50, and .75) may themselves yield different recommendations. As such we undertake a simplified analysis of the Rowen et al. sample data. For each of the 102 health states, we draw the subsample for whom that state was valued in the cTTO exercises and apply the same inclusion criteria as used in Rowen et al. The subsamples differ in size, but all contain at least 70 observations. For each subsample we compute the means (which are identical to those reported in table 2 of Rowen et al.) as well as the .50, .25, and .75 quantiles.

[8] The EQ-5D-5L is a widely used health-related quality of life (HRQOL) classification system developed by the EuroQol Foundation that is defined by five dimensions of health (mobility, self-care, usual activities, pain/discomfort, anxiety/depression) wherein each dimension is self-rated at one of five levels (Rowen et al., 2026). For mobility, self-care, and usual activities the five levels are: no problems (1), slight problems (2), moderate problems (3), severe problems (4), and unable to do (5). For pain/discomfort and anxiety/depression the five levels are: none (1), slight (2), moderate (3), severe (4), and extreme (5). A five-digit health profile summarizes a single pattern of HRQOL, e.g. 21232 signifies slight mobility problems, no self-care problems, slight usual activity problems, moderate pain/discomfort, and slight anxiety/depression.

The $u_{jk}$ were analyzed by Rowen et al. as if they are known single values, but in fact the $u_{jk}$ are more appropriately conceived as intervals. Given the cTTO scaling, $u_{jk}$ can range from –1 to 1 with $u_{jk} = 1$ indicating perfect health. Rigorous computation of each subject's interval values arising from the cTTO exercises conducted in the Rowen et al. study turns out to be quite complex (see Stolk et al., 2019; also personal communication from E. Stolk, November 10, 2026), and in any event would require information not available in the public use sample. For illustrative purposes we instead use a simpler approach that maps the point values available in the public use sample into intervals; see Appendix table A1 for the mapping we use. The full results for the point values and for the interval values described above are reported in Appendix tables A2 and A3, respectively.

Table 1 presents for illustrative purposes a subset of twenty profiles whose rows appear in descending order according to the point values' means. The table shows the point values' means (corresponding to table 2 in Rowen et al.) and the lower and upper interval bounds for the .50, .25, and .75 quantiles. With reference to the point values, these are the profiles where:

- the mean values are largest (11211, 11121, 11112, 21111, 12111, 12121), or
- the absolute difference between the point values' means and medians is largest (51451, 55225, 55333, 55424, 24443), or
- differences in the ordering of (a) means vs. medians (55333, 51152, 51124, 23152, 34342) or (b) .25-quantiles and .75-quantiles (55333, 11425, 52454, 23255, 35143) are largest.

The full results are presented in Appendix table A2.

Substantial differences between mean and median values for particular profiles indicate that utilitarian and quantile welfare criteria sometimes yield substantively different summary inferences about both the level and the ordering of population health values in these profiles. Differences across profiles in the relative magnitudes of mean and median values imply that utilitarian and .5-quantile welfare criteria sometimes provide discordant relative summaries of alternative health states. Differences between the intervals for .25-quantile and .75-quantiles indicate that the choice of a particular quantile as the basis of quantile welfare evaluation can yield different conclusions about the relative values of alternative health states. Overall, we conclude that the distinction between utilitarian and quantile welfare evaluation is not only conceptually important but (at least for these data) empirically important as well.

## 6. Conclusion

We believe that this paper's analysis should be of interest both conceptually and empirically. The conceptual interest may arise in large part because the paper's juxtaposition of utilitarian and quantile welfare may spur decision makers to assess or reassess the goals they are attempting to achieve in the populations whose welfare is of concern. An especially salient case is the UK National Health Service, which explicitly uses utilitarian welfare economics to decide whether or not to fund care with new health technologies.

Rethinking goals is vital if utilitarian and quantile welfare maximization lead to different recommendations about which policies to adopt. Unquestioningly treating utilitarian welfare maximization as the only reasonable way to evaluate policies seems to us indefensible, particularly when quantile welfare maximization offers a solidly grounded alternative. If nothing else, we hope to prod decision makers to think carefully about their welfare functions. Of course, if quantile welfare is relevant, the issue of *which* quantile(s) matter must be faced squarely. We hope the empirical analysis presented in section 5 provides compelling motivation to do this.

Much of the paper's empirical interest arises because the econometric frameworks we propose should be of direct relevance across a variety of real-world evaluation settings. We have focused on evaluations based on EQ-5D and corresponding TTO-based interval utility measures, as these have been prominent in health cost-effectiveness analysis. Our results are applicable much more generally; indeed, they are more straightforward to implement if utilities are point- rather than interval-observed. Given that regression of health outcomes on personal covariates is often of interest in applied work, future research might productively focus on econometric tools that would enable empirical researchers to straightforwardly implement nonparametric (ideally) or parametric versions of quantile interval regression, perhaps extending the work of Beresteanu and Sasaki (2021).[9]

[9] Confronted with interval outcome measures and interested in pursuing quantile regression, researchers might be tempted to consider analytical shortcuts. Perhaps most obvious would be to compute interval midpoints and treat them as data. Beresteanu and Sasaki (2021) warn against this approach. An alternative would be to consider a fully parametric strategy, estimating interval regression models akin to homoskedastic-normal Tobit models using maximum likelihood (e.g. Stata's *intreg* procedure). The resulting estimates of conditional distributions could be used to derive those distributions' conditional quantiles. This strategy is defensible if the homoskedastic-normal assumption correctly describes the data-generating process, although it effectively treats the upper bound as a right-censoring point. However, failure of a homoskedastic-normal assumption would render such estimates questionable.

Appendix: More on Issues with Use of Tobit Models

The canonical Tobit specification assumes that an underlying latent (partially observed) quality of life variable $q^*$ is distributed N(xb, v). Here x is a covariate vector that includes h and may also include individual demographic attributes, b are the corresponding coefficients, and $v > 0$ is a constant variance (i.e., the distribution of $q^*$ is homoskedastic). Tobit is implemented using data on the observed x and on the observed counterpart of $q^*$, $q = L\cdot 1(q^* \leq L) + U\cdot 1(q^* \geq U) + q^*\cdot 1(L < q^* < U)$. Thus, the observed q is interpreted to be a doubly-censored version of $q^*$, even though U is logically a true upper bound on utility and only L is a censored lower bound.

The fragility of the Tobit model, manifest in the sensitivity of estimates to departures from the homoskedastic normal assumption, has long been known. See, for example, Hurd (1979) and Arabmazur and Schmidt (1983). A question that arises sometimes in estimation of Tobit models is whether the estimands of interest to decision makers are features of the conditional distribution of latent outcomes like $E(q^*|x) = xb$, or features of the conditional distribution of observed outcomes like $E(q|x)$, which does not in general equal xb.[10] Neither is unambiguously correct or incorrect across all circumstances. Note that, if the Tobit model is specified correctly, the conditional mean $E(q^*|x)$ coincides with the conditional .5-quantile of $q^*$ (the conditional median) because N(xb, v) is symmetric around xb. In contrast, $E(q|x)$ does not generally coincide with the conditional median of q.

Another issue that should be resolved before implementing a Tobit estimation strategy is whether the bounds L and U represent censoring values or have some different meaning. Doubly-censored Tobit treats both as censoring values, suggesting that it is meaningful to conceive of values of $q^*$ in $(-\infty, L)$ and in $(U, +\infty)$. Considering the lower bound L to express censoring is defensible since some methods of utility elicitation could in principle produce values less than L. It is not clear, however, how to conceive of values $q^*$ in $(U, +\infty)$, given that U is defined as the value of full health. Reconciliation of these issues appears to us essential if one is contemplating Tobit-type estimation strategies.

[10] If a decision maker's interests exclusively concern $E(q|\mathbf{x})$, then alternatives to Tobit like fractional regression (Papke and Wooldridge, 1996) may in some instances be attractive.

Table 1: Means and Quantile Interval Bounds for Selected Profiles

| Profile | Sample Size | Mean | .50-Quantile | | .25-Quantile | | .75-Quantile | |
|---|---|---|---|---|---|---|---|---|
| | | | LB | UB | LB | UB | LB | UB |
| 11211 | 187 | 0.954 | 0.95 | 1 | 0.90 | 1 | 0.95 | 1 |
| 11121 | 295 | 0.947 | 0.95 | 1 | 0.90 | 1 | 0.95 | 1 |
| 11112 | 197 | 0.945 | 0.95 | 1 | 0.90 | 1 | 0.95 | 1 |
| 21111 | 199 | 0.942 | 0.95 | 1 | 0.90 | 1 | 0.95 | 1 |
| 12111 | 289 | 0.922 | 0.95 | 1 | 0.85 | 0.95 | 0.95 | 1 |
| 12121 | 86 | 0.912 | 0.90 | 1 | 0.85 | 0.95 | 0.95 | 1 |
| 11425 | 88 | 0.368 | 0.45 | 0.55 | 0.15 | 0.25 | 0.75 | 0.85 |
| 34342 | 84 | 0.311 | 0.45 | 0.55 | 0.25 | 0.35 | 0.65 | 0.75 |
| 23152 | 104 | 0.307 | 0.45 | 0.55 | 0.15 | 0.25 | 0.65 | 0.75 |
| 51124 | 77 | 0.305 | 0.45 | 0.55 | 0.25 | 0.35 | 0.70 | 0.80 |
| 51152 | 90 | 0.294 | 0.45 | 0.55 | 0.05 | 0.15 | 0.55 | 0.65 |
| 55333 | 88 | 0.228 | 0.45 | 0.55 | -∞ | 0 | 0.65 | 0.75 |
| 35143 | 84 | 0.228 | 0.35 | 0.45 | 0.05 | 0.15 | 0.50 | 0.60 |
| 23255 | 76 | 0.085 | 0.25 | 0.35 | -∞ | 0 | 0.55 | 0.65 |
| 24443 | 79 | 0.044 | 0.25 | 0.35 | -∞ | 0 | 0.45 | 0.55 |
| 55225 | 84 | 0.027 | 0.25 | 0.35 | -∞ | 0 | 0.45 | 0.55 |
| 51451 | 86 | 0.023 | 0.25 | 0.35 | -∞ | 0 | 0.45 | 0.55 |
| 55424 | 94 | -0.066 | 0.15 | 0.25 | -∞ | 0 | 0.35 | 0.45 |
| 52454 | 95 | -0.159 | -∞ | 0.05 | -∞ | 0 | 0.45 | 0.55 |

Table A1: Mapping from Utility Values Provided in Public Use Sample to Interval Values

| Reported Value | Coded Upper Bound | Coded Lower Bound |
|---|---|---|
| 1 | 1 | .95 |
| .95 | 1 | .90 |
| .90 | .95 | .85 |
| .85 | .90 | .80 |
| .80 | .85 | .75 |
| .75 | .80 | .70 |
| .70 | .75 | .65 |
| .65 | .70 | .60 |
| .60 | .65 | .55 |
| .55 | .60 | .50 |
| .50 | .55 | .45 |
| .45 | .50 | .40 |
| .40 | .45 | .35 |
| .35 | .40 | .30 |
| .30 | .35 | .25 |
| .25 | .30 | .20 |
| .20 | .25 | .15 |
| .15 | .20 | .10 |
| .10 | .15 | .05 |
| .05 | .10 | 0 |
| 0 | .05 | -∞ |
| Values < 0 | 0 | -∞ |

Table A2: Results for All Profiles: Point Mean Values; Lower and Upper Bounds on Quantile Values

| Profile | Sample Size | Mean | .50-Quantile | | .25-Quantile | | .75-Quantile | |
|---|---|---|---|---|---|---|---|---|
| | | | LB | UB | LB | UB | LB | UB |
| 11211 | 187 | 0.954 | 0.95 | 1 | 0.9 | 1 | 0.95 | 1 |
| 11121 | 295 | 0.947 | 0.95 | 1 | 0.9 | 1 | 0.95 | 1 |
| 11112 | 197 | 0.945 | 0.95 | 1 | 0.9 | 1 | 0.95 | 1 |
| 21111 | 199 | 0.942 | 0.95 | 1 | 0.9 | 1 | 0.95 | 1 |
| 12111 | 289 | 0.922 | 0.95 | 1 | 0.85 | 0.95 | 0.95 | 1 |
| 12121 | 86 | 0.912 | 0.9 | 1 | 0.85 | 0.95 | 0.95 | 1 |
| 11122 | 92 | 0.895 | 0.9 | 1 | 0.85 | 0.95 | 0.95 | 1 |
| 21112 | 101 | 0.887 | 0.9 | 1 | 0.85 | 0.95 | 0.95 | 1 |
| 12112 | 106 | 0.884 | 0.9 | 1 | 0.85 | 0.95 | 0.95 | 1 |
| 11212 | 101 | 0.880 | 0.9 | 1 | 0.85 | 0.95 | 0.95 | 1 |
| 13122 | 94 | 0.865 | 0.9 | 1 | 0.75 | 0.85 | 0.95 | 1 |
| 11221 | 99 | 0.863 | 0.9 | 1 | 0.8 | 0.9 | 0.95 | 1 |
| 11421 | 81 | 0.754 | 0.75 | 0.85 | 0.65 | 0.75 | 0.85 | 0.95 |
| 13313 | 87 | 0.738 | 0.75 | 0.85 | 0.65 | 0.75 | 0.85 | 0.95 |
| 25113 | 90 | 0.704 | 0.75 | 0.85 | 0.55 | 0.65 | 0.85 | 0.95 |
| 11532 | 99 | 0.684 | 0.75 | 0.85 | 0.55 | 0.65 | 0.85 | 0.95 |
| 25122 | 83 | 0.666 | 0.75 | 0.85 | 0.55 | 0.65 | 0.85 | 0.95 |
| 14113 | 91 | 0.663 | 0.75 | 0.85 | 0.55 | 0.65 | 0.85 | 0.95 |
| 25222 | 92 | 0.658 | 0.75 | 0.85 | 0.55 | 0.65 | 0.85 | 0.95 |
| 35311 | 89 | 0.630 | 0.75 | 0.85 | 0.45 | 0.55 | 0.85 | 0.95 |
| 42321 | 90 | 0.626 | 0.7 | 0.8 | 0.45 | 0.55 | 0.85 | 0.95 |
| 12513 | 89 | 0.614 | 0.65 | 0.75 | 0.45 | 0.55 | 0.75 | 0.85 |
| 25331 | 84 | 0.599 | 0.65 | 0.75 | 0.45 | 0.55 | 0.75 | 0.85 |
| 42431 | 77 | 0.587 | 0.65 | 0.75 | 0.4 | 0.5 | 0.75 | 0.85 |
| 35332 | 97 | 0.579 | 0.65 | 0.75 | 0.45 | 0.55 | 0.75 | 0.85 |
| 34232 | 91 | 0.554 | 0.65 | 0.75 | 0.4 | 0.5 | 0.75 | 0.85 |
| 53221 | 85 | 0.554 | 0.7 | 0.8 | 0.45 | 0.55 | 0.85 | 0.95 |
| 11235 | 90 | 0.499 | 0.55 | 0.65 | 0.35 | 0.45 | 0.75 | 0.85 |
| 32314 | 93 | 0.491 | 0.55 | 0.65 | 0.35 | 0.45 | 0.75 | 0.85 |
| 45133 | 73 | 0.488 | 0.55 | 0.65 | 0.35 | 0.45 | 0.75 | 0.85 |
| 52431 | 85 | 0.486 | 0.55 | 0.65 | 0.25 | 0.35 | 0.75 | 0.85 |
| 13224 | 91 | 0.485 | 0.6 | 0.7 | 0.35 | 0.45 | 0.75 | 0.85 |
| 11414 | 87 | 0.484 | 0.55 | 0.65 | 0.3 | 0.4 | 0.7 | 0.8 |
| 23514 | 82 | 0.463 | 0.55 | 0.65 | 0.25 | 0.35 | 0.7 | 0.8 |
| 12334 | 89 | 0.448 | 0.45 | 0.55 | 0.25 | 0.35 | 0.75 | 0.85 |
| 25513 | 91 | 0.446 | 0.55 | 0.65 | 0.3 | 0.4 | 0.75 | 0.85 |
| 31514 | 76 | 0.441 | 0.45 | 0.55 | 0.25 | 0.35 | 0.65 | 0.75 |
| 53412 | 83 | 0.441 | 0.55 | 0.65 | 0.25 | 0.35 | 0.75 | 0.85 |
| 21334 | 93 | 0.432 | 0.55 | 0.65 | 0.25 | 0.35 | 0.75 | 0.85 |
| 54231 | 94 | 0.431 | 0.55 | 0.65 | 0.25 | 0.35 | 0.75 | 0.85 |
| 22434 | 92 | 0.423 | 0.45 | 0.55 | 0.15 | 0.25 | 0.7 | 0.8 |
| 21315 | 89 | 0.404 | 0.5 | 0.6 | 0.25 | 0.35 | 0.75 | 0.85 |
| 45233 | 79 | 0.395 | 0.45 | 0.55 | 0.3 | 0.4 | 0.65 | 0.75 |
| 45413 | 94 | 0.378 | 0.45 | 0.55 | 0.25 | 0.35 | 0.65 | 0.75 |
| 23242 | 95 | 0.376 | 0.5 | 0.6 | 0.25 | 0.35 | 0.65 | 0.75 |
| 31524 | 86 | 0.372 | 0.45 | 0.55 | 0.2 | 0.3 | 0.65 | 0.75 |
| 12514 | 94 | 0.372 | 0.45 | 0.55 | 0.2 | 0.3 | 0.7 | 0.8 |
| 11425 | 88 | 0.368 | 0.45 | 0.55 | 0.15 | 0.25 | 0.75 | 0.85 |
| 43324 | 78 | 0.356 | 0.45 | 0.55 | 0.15 | 0.25 | 0.65 | 0.75 |
| 23415 | 82 | 0.355 | 0.45 | 0.55 | 0.15 | 0.25 | 0.65 | 0.75 |
| 52215 | 70 | 0.337 | 0.45 | 0.55 | 0.15 | 0.25 | 0.65 | 0.75 |
| 42115 | 88 | 0.334 | 0.45 | 0.55 | 0.2 | 0.3 | 0.65 | 0.75 |
| 31525 | 84 | 0.332 | 0.45 | 0.55 | 0.15 | 0.25 | 0.55 | 0.65 |
| 15151 | 89 | 0.326 | 0.45 | 0.55 | 0.15 | 0.25 | 0.65 | 0.75 |
| 55233 | 80 | 0.324 | 0.45 | 0.55 | 0.15 | 0.25 | 0.65 | 0.75 |
| 12543 | 83 | 0.319 | 0.45 | 0.55 | 0.15 | 0.25 | 0.65 | 0.75 |
| 24342 | 84 | 0.318 | 0.45 | 0.55 | 0.15 | 0.25 | 0.65 | 0.75 |
| 34342 | 84 | 0.311 | 0.45 | 0.55 | 0.25 | 0.35 | 0.65 | 0.75 |
| 23152 | 104 | 0.307 | 0.45 | 0.55 | 0.15 | 0.25 | 0.65 | 0.75 |
| 51124 | 77 | 0.305 | 0.45 | 0.55 | 0.25 | 0.35 | 0.7 | 0.8 |
| 51152 | 90 | 0.294 | 0.45 | 0.55 | 0.05 | 0.15 | 0.55 | 0.65 |
| 12244 | 86 | 0.244 | 0.35 | 0.45 | 0.1 | 0.2 | 0.55 | 0.65 |
| 43514 | 101 | 0.237 | 0.35 | 0.45 | 0.05 | 0.15 | 0.55 | 0.65 |
| 43315 | 84 | 0.235 | 0.35 | 0.45 | 0 | 0.1 | 0.6 | 0.7 |
| 21345 | 103 | 0.231 | 0.35 | 0.45 | 0.05 | 0.15 | 0.55 | 0.65 |
| 55333 | 88 | 0.228 | 0.45 | 0.55 | -∞ | 0 | 0.65 | 0.75 |
| 35143 | 84 | 0.228 | 0.35 | 0.45 | 0.05 | 0.15 | 0.5 | 0.6 |
| 32443 | 81 | 0.227 | 0.4 | 0.5 | 0 | 0.1 | 0.55 | 0.65 |
| 22144 | 94 | 0.223 | 0.35 | 0.45 | -∞ | 0.05 | 0.55 | 0.65 |
| 44342 | 79 | 0.217 | 0.4 | 0.5 | -∞ | 0.05 | 0.6 | 0.7 |
| 43542 | 74 | 0.199 | 0.35 | 0.45 | -∞ | 0.05 | 0.55 | 0.65 |
| 31443 | 95 | 0.198 | 0.35 | 0.45 | -∞ | 0.05 | 0.6 | 0.7 |
| 24543 | 87 | 0.193 | 0.35 | 0.45 | -∞ | 0.05 | 0.55 | 0.65 |
| 44125 | 91 | 0.185 | 0.35 | 0.45 | -∞ | 0.05 | 0.55 | 0.65 |
| 12344 | 91 | 0.181 | 0.25 | 0.35 | -∞ | 0.05 | 0.55 | 0.65 |
| 34515 | 92 | 0.176 | 0.25 | 0.35 | -∞ | 0 | 0.55 | 0.65 |
| 33253 | 85 | 0.174 | 0.35 | 0.45 | -∞ | 0.05 | 0.55 | 0.65 |
| 24553 | 89 | 0.144 | 0.25 | 0.35 | -∞ | 0.05 | 0.45 | 0.55 |
| 53243 | 84 | 0.136 | 0.25 | 0.35 | -∞ | 0 | 0.45 | 0.55 |
| 52335 | 84 | 0.132 | 0.2 | 0.3 | -∞ | 0 | 0.5 | 0.6 |
| 23255 | 76 | 0.085 | 0.25 | 0.35 | -∞ | 0 | 0.55 | 0.65 |
| 24443 | 79 | 0.044 | 0.25 | 0.35 | -∞ | 0 | 0.45 | 0.55 |
| 21444 | 82 | 0.034 | 0.15 | 0.25 | -∞ | 0 | 0.45 | 0.55 |
| 55225 | 84 | 0.027 | 0.25 | 0.35 | -∞ | 0 | 0.45 | 0.55 |
| 51451 | 86 | 0.023 | 0.25 | 0.35 | -∞ | 0 | 0.45 | 0.55 |
| 35245 | 89 | 0.020 | 0.15 | 0.25 | -∞ | 0 | 0.45 | 0.55 |
| 34244 | 102 | 0.015 | 0.15 | 0.25 | -∞ | 0 | 0.45 | 0.55 |
| 34155 | 79 | 0.011 | 0.2 | 0.3 | -∞ | 0 | 0.45 | 0.55 |
| 54342 | 88 | 0.005 | 0.15 | 0.25 | -∞ | 0 | 0.45 | 0.55 |
| 53244 | 81 | -0.016 | 0.15 | 0.25 | -∞ | 0 | 0.45 | 0.55 |
| 45144 | 89 | -0.028 | 0.1 | 0.2 | -∞ | 0 | 0.45 | 0.55 |
| 54144 | 92 | -0.051 | 0 | 0.1 | -∞ | 0 | 0.45 | 0.55 |
| 55424 | 94 | -0.066 | 0.15 | 0.25 | -∞ | 0 | 0.35 | 0.45 |
| 54153 | 83 | -0.069 | 0.1 | 0.2 | -∞ | 0 | 0.45 | 0.55 |
| 14554 | 91 | -0.088 | 0.05 | 0.15 | -∞ | 0 | 0.45 | 0.55 |
| 24445 | 94 | -0.120 | 0 | 0.1 | -∞ | 0 | 0.35 | 0.45 |
| 52454 | 95 | -0.159 | -∞ | 0.05 | -∞ | 0 | 0.45 | 0.55 |
| 44345 | 89 | -0.172 | -∞ | 0.05 | -∞ | 0 | 0.35 | 0.45 |
| 52455 | 86 | -0.177 | -∞ | 0.05 | -∞ | 0 | 0.25 | 0.35 |
| 44553 | 100 | -0.216 | -∞ | 0.05 | -∞ | 0 | 0.25 | 0.35 |
| 43555 | 85 | -0.256 | -∞ | 0 | -∞ | 0 | 0.25 | 0.35 |
| 55555 | 1056 | -0.480 | -∞ | 0 | -∞ | 0 | -∞ | 0.05 |